\documentclass[twocolumn,showpacs,preprintnumbers,superscriptaddress,amsmath,prl]{revtex4}
\usepackage{graphicx}
\usepackage{verbatim}

\begin{document}


\title{Room Temperature Electrical Detection of Spin Coherence in
C$_{60}$}

\author{W. Harneit}
\email[Corresponding author: ]{harneit@physik.fu-berlin.de}
\affiliation{Freie Universit\"at Berlin, Institut f\"ur
Experimentalphysik, Arnimalle 14, 14195 Berlin, Germany}
\affiliation{Hahn--Meitner--Institut Berlin, Abteilung Heterogene
Materialsysteme, Glienicker Str. 100, 14109 Berlin, Germany}

\author{C. Boehme}
\email[Corresponding author: ]{boehme@physics.utah.edu}
\affiliation{University of Utah, Physics Department, 115 S 1400 E
Suite 201, Salt Lake City, Utah 84112-0830, USA}
\affiliation{Hahn--Meitner--Institut Berlin, Abteilung
Silizium--Photovoltaik, Kekul\'{e}str. 5, 12489 Berlin, Germany}

\author{S. Schaefer}
\affiliation{Freie Universit\"at Berlin, Institut f\"ur
Experimentalphysik, Arnimalle 14, 14195 Berlin, Germany}

\author{K. Huebener}
\affiliation{Freie Universit\"at Berlin, Institut f\"ur
Experimentalphysik, Arnimalle 14, 14195 Berlin, Germany}
\affiliation{Hahn--Meitner--Institut Berlin, Abteilung Heterogene
Materialsysteme, Glienicker Str. 100, 14109 Berlin, Germany}

\author{K. Fostiropoulos}
\affiliation{Hahn--Meitner--Institut Berlin, Abteilung Heterogene
Materialsysteme, Glienicker Str. 100, 14109 Berlin, Germany}

\author{K. Lips}
\affiliation{Hahn--Meitner--Institut Berlin, Abteilung
Silizium--Photovoltaik, Kekul\'{e}str. 5, 12489 Berlin, Germany}

\date{\today}

\begin{abstract}
An experimental demonstration of electrical detection of coherent
spin motion of weakly coupled, localized electron spins in thin
Fullerene C$_{60}$ films at room temperature is presented. Pulsed
electrically detected magnetic resonance experiments on vertical
photocurrents through Al/C$_{60}$/ZnO samples showed that an
electron spin Rabi oscillation is reflected by transient current
changes. The nature of possible microscopic mechanisms responsible
for this spin to charge conversion as well as its implications for
the readout of endohedral Fullerene (N@C$_{60}$) spin qubits are
discussed.
\end{abstract}


\maketitle

The need for very sensitive spin measurement techniques for small
electron ensembles, possibly even single spins, in organic
semiconductors has grown significantly in recent years, since: (i)
The development of organic spintronic devices has progressed and
the possibility of spin-injection~\cite{Dediu02},
spin--transport~\cite{Tsukagoshi99}, and
spin--valves~\cite{Xiong04,Petta04} is increasingly investigated.
(ii) Coherent spin measurements have become important for the
investigation of spin-dependent processes in organic devices under
operating conditions (room temperature). Examples are
singlet-triplet mixing mechanisms in organic light emitting
diodes~\cite{Forrest04} or spin-dependent recombination in organic
solar cells, which are both relevant in order to assess the true
energy efficiency limitations of these devices. (iii) Organic
molecular systems such as endohedral fullerenes exhibit extremely
long spin coherence times at room temperature~\cite{Harneit02}
suggesting a possible use for room temperature quantum information
applications~\cite{Harneit02, Mehring04, Morton05, Morton06}.

Traditional electron spin measurement techniques such as electron
paramagnetic resonance (EPR) lack the sensitivity required for the
minute number of spins present in the low dimensional geometries
of many organic semiconductor devices. Several experimental
demonstrations of much more sensitive measurement approaches based
on electrical spin detection using spin to charge conversion
mechanisms have been reported for various electronic or nuclear
spin systems such as localized point defects of inorganic
semiconductors~\cite{Boehme03a, Xia04, Greentree05, Stegner06} or
semiconductor quantum dots~\cite{Petta05, Koppens06}. Most of
these approaches are confined to low temperatures ($T\leq 5\,$K),
in part due to the short coherence times of the comparatively
strongly spin-orbit-coupled spins in inorganic semiconductors at
higher temperatures. In contrast, organic semiconductors are known
to possess paramagnetic centers with extraordinarily long
coherence times at higher temperatures and even at room
temperature~\cite{Harneit02}. However, to our knowledge, there
have been no reports of experimental demonstrations for coherent
electrical spin detection in organic semiconductors so far.

In the following, we report on the experimental demonstration of
coherent spin measurements of localized paramagnetic states in
Fullerene C$_{60}$ layers at room temperature by purely electrical
means, namely through transient measurements of electric currents
that are governed by spin-selection rules. To this end, pulsed
electrically detected magnetic resonance (pEDMR) experiments were
conducted on vertical photocurrents through thin C$_{60}$ films in
transient nutation style experiments~\cite{Stegner06}.

Several samples were prepared for this study, with $80-300\,$nm
thick Fullerene films sandwiched between two Al electrodes, or
between Al and ZnO. The influence of these variations on the
results is minor and will be addressed elsewhere. All layers were
evaporated in ultra--high vacuum and the samples were encapsulated
in quartz tubes in an inert--gas glove box without intermediate
exposure to air that could negatively affect the electronic
properties of the thin films. The active device area was
$2\,\mathrm{mm}\times 2\,\mathrm{mm}$ and thin contact stripes
were extended on a $2.6\,\mathrm{mm}\times 58\,\mathrm{mm}$
substrate towards a region outside the resonator, in order to
minimize field mode distortions in the resonator. The pEDMR
measurements were carried out on a commercial pulsed ESR X--band
spectrometer (Bruker E580). The sample was illuminated through the
optical view port of the resonator and through one of the contact
layers with a halogen lamp ($110\,$W) and an argon ion laser
($488\,$nm, up to $1\,$W power). The photo--induced bias current
$I_0$ was stabilized by a current source with a long time
constant, so that the transient current response (extending to
$\approx 100\,\mu$s) after the microwave pulses was not averaged
out. The amplitude $B_1$ of the microwave field was calibrated
using a 4-hydroxy-2,2,6,6-tetramethylpiperidine-N-oxyl (Tempol)
standard (Sigma Aldrich).

\begin{figure}
\centering
\includegraphics[width=\columnwidth]{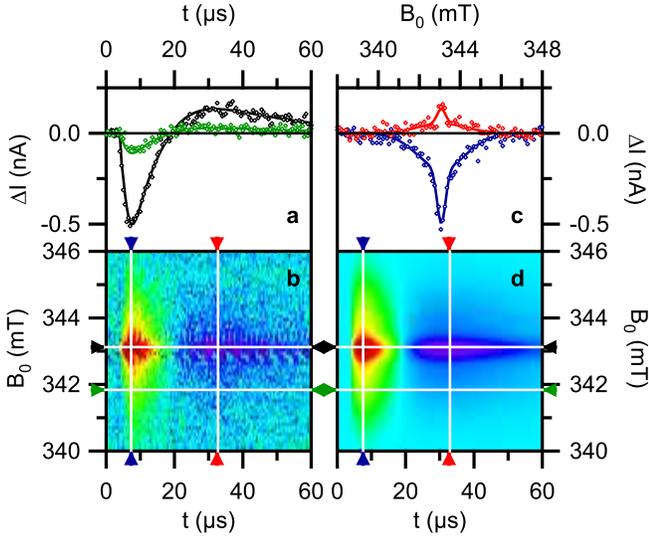}
\caption{(online color) Transient current $\Delta I$ of a C$_{60}$
film illuminated with a halogen lamp ($\approx 10\,$mW/cm$^2$,
$I_0 = 10\,\mu$A) after $320\,$ns long microwave pulses ($B_1
\approx 63\,\mu$T). Time traces (a) were recorded at different
magnetic fields $B_0$ to afford the 2D dataset (b). 1D cuts along
the field dimension (c) reveal the spectral behavior. Panel (d)
displays a global fit of the data in (b) together with an
indication where the cuts (a, c) were taken; in these, the data
are represented by markers, whereas the global fit is drawn as
lines.} \label{FigTransient}
\end{figure}

Figure \ref{FigTransient} shows the transient current response
$\Delta I$ to microwave pulses as a function of (a) time $t$ and
(c) magnetic field $B_0$. The 2D dataset (b) was modelled with a
global 2D fit (d). The fit function $\Delta I(t,B_0) =
I_\mathrm{ph}(t)\Delta(B_0)$ is a product function of the time
evolution of the current
\begin{equation}%
I_\mathrm{ph}(t)=\left(1-e^{-\frac{t}{t_\mathrm{s}}}\right)
\sum_{j=1..3}I_j\,e^{-\frac{t-t_\mathrm{d}}{\tau_j}}
\end{equation}%
and the spectroscopic function
\begin{equation}%
\Delta\left(B_0\right) = \frac{2}{\pi} \sum_{k} \frac{\Delta
B_k}{4\left(B_0 - \frac{h\nu_0}{\mu_\mathrm{B}}g_k^{-1}\right)^2
+\Delta B_k^2}%
\end{equation}%
which is a multiple Lorentzian normalized to unit area containing
Planck's constant $h$, Bohr's magneton $\mu_\mathrm{B}$, the
microwave frequency $\nu_0$, and the Land\'{e} factor $g_k$ and
full width at half height $\Delta B_k$ for each resonance line
$k$. The time evolution is described by the signed weights $I_j$
and time constants $\tau_j$ of the multi exponential decay
expected from theory~\cite{Boehme03b}, a pulse trigger delay
$t_\mathrm{d}$, and a saturation time $t_\mathrm{s}$ describing
the response of the detection circuit which is due to a finite
rise time.

Due to the product decomposition of $\Delta I(t,B_0)$, the
spectroscopic information can be interpreted independently of the
time evolution of the current if the latter is integrated out. The
resulting charge--equivalent pEDMR signal $Q(B_0) = \int\Delta
I(t,B_0) dt$ contains all spectroscopic information. In general,
the integration limits are chosen to give maximal $Q$ and the
integration can be carried out using the hardware of the ESR
spectrometer. Careful analysis reveals that the spectroscopic
information contained in $Q(B_0)$ does not depend on that choice.
The magnitude of $Q$ is a direct measure of the number of carriers
involved in the spin--dependent process. It turns out that our
room temperature pEDMR signal is due to $\approx 1-2\times10^4$
elementary charges. In this small--ensemble regime, an exact model
of the charge carrier dynamics is needed in order to relate the
number of charges to the number of spins since the constants $I_j$
and $\tau_j$ depend in general on numerous parameters such as
capture and emission cross sections of traps as well as carrier
generation and recombination rates. Even in the best of cases,
these parameters are hard to deconvolve
quantitatively~\cite{Boehme04}, and we do not attempt this in the
present paper. The spectroscopic information obtained from the
data of Fig.~\ref{FigTransient} and similar data (not shown here)
can be summarized as follows: (i) The signal is observed on some,
but not on all samples prepared. (ii) It is observed only under
illumination, and consistently absent in the dark. (iii) The
initial sign of the signal is negative, corresponding to current
quenching while at times $t\gtrsim 20\,\mu$s, the sign reverses to
current enhancement. (iv) The signal magnitude $|\Delta I|$
depends on the samples investigated (by a factor of two at least),
but (v) $|\Delta I|$ depends only weakly on the illumination
intensity or wavelength. (vi) The resonance condition is fulfilled
at $g = 2.0018(5)$ and (vii) the smallest observed line width is
$\Delta B_a\leq 0.3\,$mT. (viii) Measurements at low amplitude
$B_1<0.2\,$mT show a second, much broader line ($\Delta B_b\approx
3\,$mT) at the same $g$ factor within our resolution.

\begin{figure}
\centering
\includegraphics[width=\columnwidth]{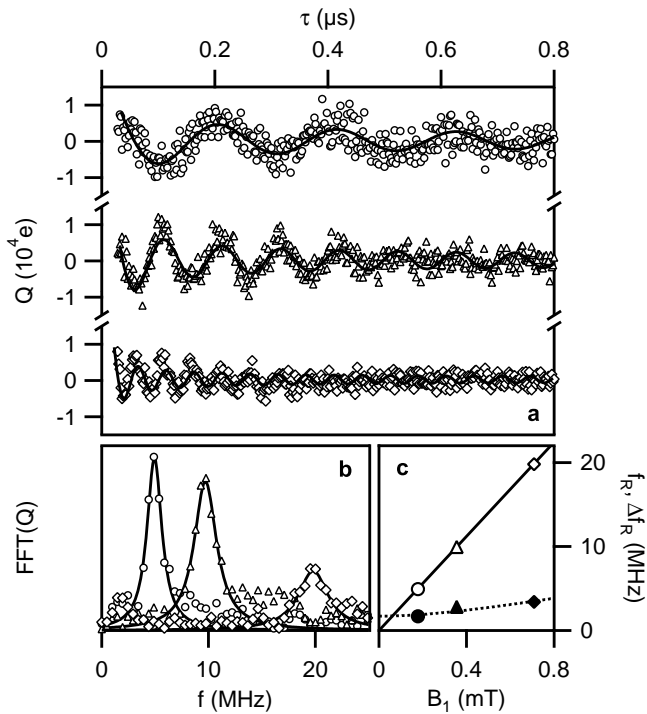}
\caption{(a) Oscillatory part of the observable $Q$ as a function
of microwave pulse length at different pulse amplitudes $B_1$ for
a C$_{60}$ film illuminated with an argon ion laser
($<100\,$mW/cm$^2$, $I_0 = 50\,\mu$A). The lines are fits using
the theoretical curve $T(\alpha=\gamma B_1\tau)$ given in
Eq.~(\ref{Eq1}). (b) Fast Fourier transforms of the data in (a)
and Lorentzian fits. (c) Rabi frequencies $f_\mathrm{R}$ (open
symbols) and resonance widths $\Delta f_\mathrm{R}$ (closed
symbols) extracted from the fits in (b). The full line in (c) is
$f_\mathrm{R}=\gamma B_1/2\pi$, the expected behavior of a spin $S
= \frac12$. The dotted line is the prediction of the line width as
a function of $B_1$ based on the assumption of an
$\approx\,15\,$\% field inhomogeneity.} \label{FigRabi}
\end{figure}

Figure \ref{FigRabi} shows the evolution of $Q$ as a function of
microwave pulse length $\tau$ for three different microwave
amplitudes $B_1$. The observed oscillatory behavior is due to
transient nutation of the spins. The coherent pulses induce Rabi
oscillations of angular frequency $\omega_\mathrm{R} =
\kappa_S\omega_1 = \kappa_S\gamma B_1$, where $\gamma= g
\mu_\mathrm{B} / \hbar$ is the gyromagnetic ratio. The Rabi
frequency $\omega_\mathrm{R}$ is indicative of the spin quantum
number $S$ via the factor $\kappa_S$~\cite{Boehme03b}, expected to
be $\kappa_S = 1$ for a doublet $S = \frac12$, $\kappa_S =
\sqrt{2}$ for a pure triplet $S = 1$, and $\kappa_S = 2$ for a
pair of coupled $S = \frac12$ spins when their spectroscopic
separation is small: $\Delta\omega_\mathrm{L} = |g_1-g_2|/g_0
\,\omega_0\ll\omega_1$. As demonstrated in Fig.~\ref{FigRabi}a by
the comparison of the data to the "transient function"
$T(\alpha)$~\cite{Boehme03b}, we observe a Rabi frequency
$\omega_\mathrm{R} = \omega_1=\gamma B_1$, as expected for $S =
\frac12$. It can easily be shown that $T(\alpha)$ is related to
the Bessel function of the first kind $J_0$,
\begin{equation}\label{Eq1}
T(\alpha)=\pi\int_0^\alpha J_0(x)dx
\end{equation}
Without its initial part $(\alpha<\pi/4)$, $T(\alpha)$ is a quasi
$2\pi$ periodic function of the nutation or "turning" angle
$\alpha=\omega_\mathrm{R}\tau$ resembling a damped sinusoid. This
"built--in" damping is due to the fact that selective excitation
of an inhomogeneously broadened line was assumed for the
derivation of $T(\alpha)$ and is not indicative of spin relaxation
in itself, it reflects a coherent dephasing process amenable to
refocussing techniques~\cite{Stoll98}. Figure~\ref{FigRabi}b shows
the fast Fourier transform of the data displayed in
Fig.~\ref{FigRabi}a, together with Lorentzian fits. \emph{A
priori}, the width of the Lorentzian $\Delta f_\mathrm{R}$ should
not depend on the microwave field strength $B_1$. The observed
slight increase in width, reported together with the Rabi
frequency in Fig.~\ref{FigRabi}c, may be due to an inhomogeneity
in the $B_1$ microwave field. In a simple model, the detected FFT
width can be described by $\Delta f_\mathrm{R}^2 = \Delta
f_\mathrm{int}^2 + (\Delta\omega_1/2\pi)^2$ where $\Delta
f_\mathrm{int}\approx1.7\,$MHz is due to the built--in damping
and a relative $B_1$ field inhomogeneity which turns out to be
$\Delta\omega_1/\omega_1\approx 15\,\%$ for the data presented in
Figure~\ref{FigRabi}c. To summarize the coherent properties at
room temperature, Rabi oscillation experiments similar to those
reported in Fig.~\ref{FigRabi} reveal (i) a $B_1$ dependence of
the Rabi frequency consistent with $S = \frac12$ character of the
signal, (ii) a damping behavior consistent with an inhomogeneously
broadened "partner line", which is visible in the spectra only at
low $B_1$ amplitudes, and (iii) coherent oscillations with a
damping time $\Delta f_\mathrm{int}^{-1} \approx 0.6\,\mu$s, which
is limited by the inhomogeneity of the spin ensemble and not by
the intrinsic $T_2$ decoherence time. An estimate for $T_2$ may be
obtained using Rabi echoes and other refocussing
techniques~\cite{Stoll98}.

The experimental data presented in Figs.~1 and 2 show that the
magnetic resonant current imprint of the paramagnetic centers with
Land\'e factor around $g\approx 2.0018(5)$ and with weakly coupled
($S=\frac{1}{2}$) systems can be observed by means of transient
measurements of the photocurrent changes after coherent spin
excitations at room temperature. The magnetic field spectra of the
current response suggest that two different resonances are present
with linewidths of $\approx 3\,$mT and $\approx 0.3\,$mT
respectively. With the given accuracies, an undisputable
association of the Land\'e factors with particular spin systems is
difficult since various paramagnetic states in C$_{60}$ are
known~\cite{Reed00}, such as the corresponding Fullerene radical
states and other impurity states, as they may occur in the used
device system. With the given experimental data, it is possible
that the current is governed by spin--dependent recombination of
excess charge carriers similar to the model described for
inorganic semiconductors~\cite{Boehme03b} assuming that the
observed spins are due to localized defect levels which enhance
recombination. It is also possible that the currents are
controlled by polaron pair quenching upon polaron pair encounter
with radicals. The latter has been referred to as the
radical--triplet pair mechanism in the literature~\cite{Steren95}.
Thus, while the experimental data presented here cannot
conclusively rule out any of these different explanations for the
observed signals, there is strong evidence for a spin--dependent
recombination mechanism of weakly coupled spin pairs ubiquitous in
Fullerenes~\cite{Reed00} since: (i) The nature and spectroscopic
properties of these pairs are in accordance with the two observed
resonances. (ii) The mechanism requires the detection of spin
$S=\frac{1}{2}$ pair partners which have been confirmed by the
data. (iii) The theoretically well investigated dynamical nature
of such pairs after coherent spin
excitations~\cite{Eickelkamp98,Boehme03b,Mehring04} is entirely
consistent with the current transients observed here. The well
understood current quenching followed by current enhancement is
characteristic for such mechanisms: It is due to the electronic
relaxation of first singlet and then triplet pair densities. In
spite of this strong agreement, we want to point out nevertheless
that in absence of theoretical predictions for pEDMR transients
due to the other spin--dependent mechanisms described in the
literature, the microscopic nature of the observed signals cannot
be verified with certainty.

\begin{figure}
\centering
\includegraphics[width=\columnwidth]{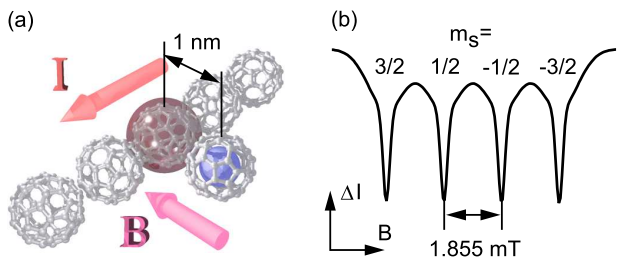}
\caption{(online color) (a) Schematic illustration of an
electrical read--out of a N@C$_{60}$ endohedral Fullerene spin.
The localized spin involved in spin-dependent transitions, marked
by a surrounding sphere, is dipolarly coupled to the adjacent
N@C$_{60}$ spin. (b) The expected spectrum for $52\,$MHz strong
coupling of the $S = \frac32$ N@C$_{60}$ spin to an $S=\frac12$
state in C$_{60}$. For details, see text.} \label{FigCoupling}
\end{figure}

The process responsible for the signals discussed above represents
a spin to charge conversion for weakly coupled, localized spin
states in a C$_{60}$ solid at room temperature. We believe that
this finding may have important implications for the development
of highly sensitive spin quantum readout schemes for endohedral
Fullerene (N@C$_{60}$) based electron spin
qubits~\cite{Harneit02}. For spin--dependent recombination through
weakly coupled spin pairs as assumed above, an identification of
the integrated current (corresponding to the number of detected
charge carriers and therefore the number of detected
spin--dependent transitions) with the number of paramagnetic
centers involved in these transitions is
possible~\cite{Boehme03b}. Hence, for the macroscopic samples used
in this study, a sensitivity of $\approx 10^4$ spins is given.
This is many orders of magnitude more sensitive than conventional
EPR measurements on C$_{60}$ and endohedral N@C$_{60}$ molecules.
Note that the given sensitivity limitations were caused by shot
noise due to spin--independent shunt currents. It is conceivable
that the same signals reach even higher sensitivities by improved
sample designs oppressing these shunt currents. Thus, we believe
the spin to charge conversion mechanism observed here may be
suitable for a dipolar coupling mediated readout of electron spins
of endohedral Fullerenes as illustrated in
Figure~\ref{FigCoupling}. Here, the localized paramagnetic
state whose spin governs charge carrier recombination %
is represented by a closed sphere due to its unknown nature. If
this probe spin is in spatial proximity to an N@C$_{60}$ qubit
molecule, the state of the probe spin will be shifted by the four
eigenstates of the $S=\frac{3}{2}$ qubit. For the case of maximal
dipolar coupling (reached at an angle $\theta=0$ between the
qubit--probe spin axis and the direction of an applied magnetic
field) and an assumed distance of $r=1\,$nm, a splitting of the
probe spin resonance into four lines with $1.855\,$mT separation
can be expected. Hence, if we assume a line width of $0.3\,$mT
(see Fig.~\ref{FigTransient}) for the electrically read probe
spin, the given line separation should be sufficient for a
distinction of the qubit eigenstates required for a readout.

In conclusion, we report on a very sensitive electrical detection
of coherent spin motion of localized, weakly coupled electron
spins in thin C$_{60}$ films at room temperature which has been
demonstrated by means of transient photocurrent measurements after
short coherent electron spin resonant excitation. The observed
data reveals the presence of electron spin Rabi oscillation of
paramagnetic centers at a Land\'e factor of $g\approx 2.0018(5)$.
The nature of the observed signals is in agreement with described
spin--dependent charge carrier pair recombination models. It has
been pointed out that the ability to coherently measure (read)
localized electron spin states in C$_{60}$ solids with high
sensitivity may be utilized as a potential readout mechanism for
endohedral N@C$_{60}$ Fullerene qubits.

We thank Jan Behrends (\emph{Hahn--Meitner--Institut}) for support
in the pEDMR measurements and Carola Meyer
(\emph{Forschungszentrum J{\"u}lich}) for helpful discussions.
This work was funded in part by the \emph{Bundesministerium
f{\"u}r Bildung und Forschung} (contract no. 03N8709).

\bibliographystyle{apsrev}

\begin{thebibliography}{23}
\expandafter\ifx\csname
natexlab\endcsname\relax\def\natexlab#1{#1}\fi
\expandafter\ifx\csname bibnamefont\endcsname\relax
  \def\bibnamefont#1{#1}\fi
\expandafter\ifx\csname bibfnamefont\endcsname\relax
  \def\bibfnamefont#1{#1}\fi
\expandafter\ifx\csname citenamefont\endcsname\relax
  \def\citenamefont#1{#1}\fi
\expandafter\ifx\csname url\endcsname\relax
  \def\url#1{\texttt{#1}}\fi
\expandafter\ifx\csname
urlprefix\endcsname\relax\def\urlprefix{URL }\fi
\providecommand{\bibinfo}[2]{#2}
\providecommand{\eprint}[2][]{\url{#2}}

\bibitem[{\citenamefont{Dediu et~al.}(2002)\citenamefont{Dediu, Murgia,
  Matacotta, Taliani, and Barbanera}}]{Dediu02}
\bibinfo{author}{\bibfnamefont{V.}~\bibnamefont{Dediu}},
  \bibinfo{author}{\bibfnamefont{M.}~\bibnamefont{Murgia}},
  \bibinfo{author}{\bibfnamefont{F.~C.} \bibnamefont{Matacotta}},
  \bibinfo{author}{\bibfnamefont{C.}~\bibnamefont{Taliani}}, \bibnamefont{and}
  \bibinfo{author}{\bibfnamefont{S.}~\bibnamefont{Barbanera}},
  \bibinfo{journal}{Solid State Commun.} \textbf{\bibinfo{volume}{122}},
  \bibinfo{pages}{181} (\bibinfo{year}{2002}).

\bibitem[{\citenamefont{Tsukagoshi et~al.}(1999)\citenamefont{Tsukagoshi,
  Alphenaar, and Ago}}]{Tsukagoshi99}
\bibinfo{author}{\bibfnamefont{K.}~\bibnamefont{Tsukagoshi}},
  \bibinfo{author}{\bibfnamefont{B.~W.} \bibnamefont{Alphenaar}},
  \bibnamefont{and} \bibinfo{author}{\bibfnamefont{H.}~\bibnamefont{Ago}},
  \bibinfo{journal}{Nature (London)} \textbf{\bibinfo{volume}{401}},
  \bibinfo{pages}{572} (\bibinfo{year}{1999}).

\bibitem[{\citenamefont{Xiong et~al.}(2004)\citenamefont{Xiong, Wu, Vardeny,
  and Shi}}]{Xiong04}
\bibinfo{author}{\bibfnamefont{Z.~H.} \bibnamefont{Xiong}},
  \bibinfo{author}{\bibfnamefont{D.}~\bibnamefont{Wu}},
  \bibinfo{author}{\bibfnamefont{Z.~V.} \bibnamefont{Vardeny}},
  \bibnamefont{and} \bibinfo{author}{\bibfnamefont{J.}~\bibnamefont{Shi}},
  \bibinfo{journal}{Nature (London)} \textbf{\bibinfo{volume}{427}},
  \bibinfo{pages}{821} (\bibinfo{year}{2004}).

\bibitem[{\citenamefont{Petta et~al.}(2004)\citenamefont{Petta, Slater, and
  Ralph}}]{Petta04}
\bibinfo{author}{\bibfnamefont{J.~R.} \bibnamefont{Petta}},
  \bibinfo{author}{\bibfnamefont{S.~K.} \bibnamefont{Slater}},
  \bibnamefont{and} \bibinfo{author}{\bibfnamefont{D.~C.} \bibnamefont{Ralph}},
  \bibinfo{journal}{Phys.\ Rev.\ Lett.} \textbf{\bibinfo{volume}{93}},
  \bibinfo{pages}{136601} (\bibinfo{year}{2004}).

\bibitem[{\citenamefont{Forrest}(2002)}]{Forrest04}
\bibinfo{author}{\bibfnamefont{S.~R.}~\bibnamefont{Forrest}},
  \bibinfo{journal}{Nature (London)} \textbf{\bibinfo{volume}{428}},
  \bibinfo{pages}{911} (\bibinfo{year}{2002}).

\bibitem[{\citenamefont{Harneit}(2002)}]{Harneit02}
\bibinfo{author}{\bibfnamefont{W.}~\bibnamefont{Harneit}},
  \bibinfo{journal}{Phys.\ Rev.\ A} \textbf{\bibinfo{volume}{65}},
  \bibinfo{pages}{032322} (\bibinfo{year}{2002}).

\bibitem[{\citenamefont{Mehring et~al.}(2004)\citenamefont{Mehring, Scherer,
  and Weidinger}}]{Mehring04}
\bibinfo{author}{\bibfnamefont{M.}~\bibnamefont{Mehring}},
  \bibinfo{author}{\bibfnamefont{W.}~\bibnamefont{Scherer}}, \bibnamefont{and}
  \bibinfo{author}{\bibfnamefont{A.}~\bibnamefont{Weidinger}},
  \bibinfo{journal}{Phys.\ Rev.\ Lett.} \textbf{\bibinfo{volume}{93}},
  \bibinfo{pages}{206603} (\bibinfo{year}{2004}).

\bibitem[{\citenamefont{Morton et~al.}(2005)\citenamefont{Morton, Tyryshkin,
  Ardavan, Porfyrakis, Lyon, and Briggs}}]{Morton05}
\bibinfo{author}{\bibfnamefont{J.~J.~L.} \bibnamefont{Morton}},
  \bibinfo{author}{\bibfnamefont{A.~M.} \bibnamefont{Tyryshkin}},
  \bibinfo{author}{\bibfnamefont{A.}~\bibnamefont{Ardavan}},
  \bibinfo{author}{\bibfnamefont{K.}~\bibnamefont{Porfyrakis}},
  \bibinfo{author}{\bibfnamefont{S.~A.} \bibnamefont{Lyon}}, \bibnamefont{and}
  \bibinfo{author}{\bibfnamefont{G.~Andrew D.} \bibnamefont{Briggs}},
  \bibinfo{journal}{Phys.\ Rev.\ Lett.} \textbf{\bibinfo{volume}{95}},
  \bibinfo{pages}{200501} (\bibinfo{year}{2005}).

\bibitem[{\citenamefont{Morton et~al.}(2006)\citenamefont{Morton, Tyryshkin,
  Ardavan, Benjamin, Porfyrakis, Lyon, and Briggs}}]{Morton06}
\bibinfo{author}{\bibfnamefont{J.~J.~L.} \bibnamefont{Morton}},
  \bibinfo{author}{\bibfnamefont{A.~M.} \bibnamefont{Tyryshkin}},
  \bibinfo{author}{\bibfnamefont{A.}~\bibnamefont{Ardavan}},
  \bibinfo{author}{\bibfnamefont{S.~C.} \bibnamefont{Benjamin}},
  \bibinfo{author}{\bibfnamefont{K.}~\bibnamefont{Porfyrakis}},
  \bibinfo{author}{\bibfnamefont{S.~A.} \bibnamefont{Lyon}}, \bibnamefont{and}
  \bibinfo{author}{\bibfnamefont{G.~Andrew D.} \bibnamefont{Briggs}},
  \bibinfo{journal}{Nature Physics} \textbf{\bibinfo{volume}{2}},
  \bibinfo{pages}{40} (\bibinfo{year}{2006}).

\bibitem[{\citenamefont{Boehme and Lips}(2003{\natexlab{a}})}]{Boehme03a}
\bibinfo{author}{\bibfnamefont{C.}~\bibnamefont{Boehme}} \bibnamefont{and}
  \bibinfo{author}{\bibfnamefont{K.}~\bibnamefont{Lips}},
  \bibinfo{journal}{Phys.\ Rev.\ Lett.} \textbf{\bibinfo{volume}{91}},
  \bibinfo{pages}{246603} (\bibinfo{year}{2003}{\natexlab{a}}).

\bibitem[{\citenamefont{Xia et~al.}(2004)\citenamefont{Xia, Martin,
  Yablonovitch, and Jiang}}]{Xia04}
  \bibinfo{author}{\bibfnamefont{M.}~\bibnamefont{Xia}},
  \bibinfo{author}{\bibfnamefont{I.}~\bibnamefont{Martin}},
  \bibinfo{author}{\bibfnamefont{E.}~\bibnamefont{Yablonovitch}}, \bibnamefont{and}
  \bibinfo{author}{\bibfnamefont{H.~W.} \bibnamefont{Jiang}},
  \bibinfo{journal}{Nature (London)} \textbf{\bibinfo{volume}{430}},
  \bibinfo{pages}{435} (\bibinfo{year}{2004}).

\bibitem[{\citenamefont{Greentree et~al.}(2005)\citenamefont{Greentree, Hamilton,
  Hollenberg, and Clark}}]{Greentree05}
  \bibinfo{author}{\bibfnamefont{A.~D.}~\bibnamefont{Greentree}},
  \bibinfo{author}{\bibfnamefont{A.~R.}~\bibnamefont{Hamilton}},
  \bibinfo{author}{\bibfnamefont{L.~C.~L.}~\bibnamefont{Hollenberg}}, \bibnamefont{and}
  \bibinfo{author}{\bibfnamefont{R.~G.} \bibnamefont{Clark}},
  \bibinfo{journal}{Phys.\ Rev.\ B} \textbf{\bibinfo{volume}{71}},
  \bibinfo{pages}{113310} (\bibinfo{year}{2005}).

\bibitem[{\citenamefont{Stegner et~al.}(2006)\citenamefont{Stegner, Boehme,
  Huebl, Stutzmann, Lips, and Brandt}}]{Stegner06}
\bibinfo{author}{\bibfnamefont{A.~R.} \bibnamefont{Stegner}},
  \bibinfo{author}{\bibfnamefont{C.}~\bibnamefont{Boehme}},
  \bibinfo{author}{\bibfnamefont{H.}~\bibnamefont{Huebl}},
  \bibinfo{author}{\bibfnamefont{M.}~\bibnamefont{Stutzmann}},
  \bibinfo{author}{\bibfnamefont{K.}~\bibnamefont{Lips}}, \bibnamefont{and}
  \bibinfo{author}{\bibfnamefont{M.~S.} \bibnamefont{Brandt}},
  \bibinfo{journal}{Nature Physics} \textbf{\bibinfo{volume}{2}},
  \bibinfo{pages}{835} (\bibinfo{year}{2006}).

\bibitem[{\citenamefont{Petta et~al.}(2005)\citenamefont{Petta, Johnson,
  Taylor, Laird, Yacoby, Lukin, Marcus, Hanson, and Gossard}}]{Petta05}
\bibinfo{author}{\bibfnamefont{J.~R.} \bibnamefont{Petta}},
  \bibinfo{author}{\bibfnamefont{A.~C.}~\bibnamefont{Johnson}},
  \bibinfo{author}{\bibfnamefont{J.~M.}~\bibnamefont{Taylor}},
  \bibinfo{author}{\bibfnamefont{E.~A.}~\bibnamefont{Laird}},
  \bibinfo{author}{\bibfnamefont{A.}~\bibnamefont{Yacoby}},
  \bibinfo{author}{\bibfnamefont{M.~D.}~\bibnamefont{Lukin}},
  \bibinfo{author}{\bibfnamefont{C.~M.}~\bibnamefont{Marcus}},
  \bibinfo{author}{\bibfnamefont{M.~P.}~\bibnamefont{Hanson}}, \bibnamefont{and}
  \bibinfo{author}{\bibfnamefont{A.~C.} \bibnamefont{Gossard}},
  \bibinfo{journal}{Science} \textbf{\bibinfo{volume}{309}},
  \bibinfo{pages}{2180} (\bibinfo{year}{2005}).

\bibitem[{\citenamefont{Koppens et~al.}(2006)\citenamefont{Koppens, Buizert,
  Tielrooij, Vink, Nowack, Meunier, Kouwenhoven, and Vandersypen}}]{Koppens06}
\bibinfo{author}{\bibfnamefont{F.~H.~L.} \bibnamefont{Koppens}},
  \bibinfo{author}{\bibfnamefont{C.}~\bibnamefont{Buizert}},
  \bibinfo{author}{\bibfnamefont{K.~J.}~\bibnamefont{Tielrooij}},
  \bibinfo{author}{\bibfnamefont{I.~T.}~\bibnamefont{Vink}},
  \bibinfo{author}{\bibfnamefont{K.~C.}~\bibnamefont{Nowack}},
  \bibinfo{author}{\bibfnamefont{T.}~\bibnamefont{Meunier}},
  \bibinfo{author}{\bibfnamefont{L.~P.}~\bibnamefont{Kouwenhoven}}, \bibnamefont{and}
  \bibinfo{author}{\bibfnamefont{L.~M.~K.} \bibnamefont{Vandersypen}},
  \bibinfo{journal}{Nature (London)} \textbf{\bibinfo{volume}{442}},
  \bibinfo{pages}{766} (\bibinfo{year}{2006}).

\bibitem[{\citenamefont{Boehme and Lips}(2003{\natexlab{b}})}]{Boehme03b}
\bibinfo{author}{\bibfnamefont{C.}~\bibnamefont{Boehme}} \bibnamefont{and}
  \bibinfo{author}{\bibfnamefont{K.}~\bibnamefont{Lips}},
  \bibinfo{journal}{Phys.\ Rev.\ B} \textbf{\bibinfo{volume}{68}},
  \bibinfo{pages}{245105} (\bibinfo{year}{2003}{\natexlab{b}}).

\bibitem[{\citenamefont{Boehme and Lips}(2004)}]{Boehme04}
\bibinfo{author}{\bibfnamefont{C.}~\bibnamefont{Boehme}} \bibnamefont{and}
  \bibinfo{author}{\bibfnamefont{K.}~\bibnamefont{Lips}},
  \bibinfo{journal}{phys.\ stat.\ sol. (c)} \textbf{\bibinfo{volume}{1}},
  \bibinfo{pages}{1255} (\bibinfo{year}{2004}).

\bibitem[{\citenamefont{Stoll et~al.}(1998)\citenamefont{Stoll, Jeschke,
  Willer, and Schweiger}}]{Stoll98}
\bibinfo{author}{\bibfnamefont{S.}~\bibnamefont{Stoll}},
  \bibinfo{author}{\bibfnamefont{G.}~\bibnamefont{Jeschke}},
  \bibinfo{author}{\bibfnamefont{M.}~\bibnamefont{Willer}}, \bibnamefont{and}
  \bibinfo{author}{\bibfnamefont{A.}~\bibnamefont{Schweiger}},
  \bibinfo{journal}{J.\ Magn.\ Reson.} \textbf{\bibinfo{volume}{130}},
  \bibinfo{pages}{86} (\bibinfo{year}{1998}).

\bibitem[{\citenamefont{Reed and Bolskar}(2000)}]{Reed00}
\bibinfo{author}{\bibfnamefont{C.~A.} \bibnamefont{Reed}} \bibnamefont{and}
  \bibinfo{author}{\bibfnamefont{R.~D.} \bibnamefont{Bolskar}},
  \bibinfo{journal}{Chem.\ Rev.} \textbf{\bibinfo{volume}{100}},
  \bibinfo{pages}{1075} (\bibinfo{year}{2000}).

\bibitem[{\citenamefont{Steren et~al.}(1995)\citenamefont{Steren, van Willigen,
  and Fanciulli}}]{Steren95}
\bibinfo{author}{\bibfnamefont{C.~A.} \bibnamefont{Steren}},
  \bibinfo{author}{\bibfnamefont{H.}~\bibnamefont{van Willigen}},
  \bibnamefont{and}
  \bibinfo{author}{\bibfnamefont{M.}~\bibnamefont{Fanciulli}},
  \bibinfo{journal}{Chem.\ Phys.\ Lett.} \textbf{\bibinfo{volume}{245}},
  \bibinfo{pages}{244} (\bibinfo{year}{1995}).

\bibitem[{\citenamefont{Eickelkamp et~al.}(1998)\citenamefont{Eickelkamp, Roth,
  and Mehring}}]{Eickelkamp98}
\bibinfo{author}{\bibfnamefont{T.}~\bibnamefont{Eickelkamp}},
  \bibinfo{author}{\bibfnamefont{S.}~\bibnamefont{Roth}}, \bibnamefont{and}
  \bibinfo{author}{\bibfnamefont{M.}~\bibnamefont{Mehring}},
  \bibinfo{journal}{Mol.\ Phys.} \textbf{\bibinfo{volume}{95}},
  \bibinfo{pages}{967} (\bibinfo{year}{1998}).

\end{thebibliography}

\end{document}